\documentclass[aps,prl,twocolumn,float]{revtex4}
\usepackage{amsmath,bm,epsfig}
\let \nn  \nonumber

\let\*\cdot
\def\<{\left\langle} \def\>{\right\rangle} \def\({\left(} \def\){\right)}
 \let\~\widetilde \let\^\widehat 

\newcommand{\C}[1]{{\mathcal{#1}}}    

\def\be{\begin{equation}}\def\ee{\end{equation}}
\def\bea{\begin{eqnarray}}\def\eea{\end{eqnarray}}
\def\bse{\begin{subequations}}\def\ese{\end{subequations}}
\newcommand{\BE}[1]{\begin{equation}\label{#1}}
\newcommand{\BEA}[1]{\begin{eqnarray}\label{#1}}
\newcommand{\BSE}[1]{\begin{subequations}\label{#1}}

\newcommand{\Eqs}[1]{Eqs.~(\ref{#1})}

\let \nn  \nonumber
\usepackage{pstricks}
\usepackage{pst-node}
\usepackage[ansinew]{inputenc}
\usepackage{amssymb,amsmath}

\def\BSE{\begin{subequations}}\def\ESE{\end{subequations}}
\let \= \equiv

\def\g{\gamma}

\def\o{\omega}

\def\be{\begin{equation}}       \def\ba{\begin{array}}

\def\ee{\end{equation}}         \def\ea{\end{array}}

\def\bea {\begin{eqnarray}}      \def\eea {\end{eqnarray}}

\def\bean{\begin{eqnarray*}}    \def\eean{\end{eqnarray*}}

\def\eps{\varepsilon}

\def\RA {\ \Rightarrow\ }

\def\<{\langle} \def\({\left(}  \def\>{\rangle} \def\){\right)}

\newtheorem{exi}{Example}

\begin{document}

\title{Dynamical cascade generation as a basic mechanism of Benjamin-Feir instability}
\author{Elena Kartashova$^{\dag}$ and Igor V. Shugan$^\ddag$}
 \email{Elena.Kartaschova@jku.at, ishugan@rambler.ru}
  \affiliation{$^\dag$ Institute for Analysis, J. Kepler University, Linz, Austria\\
$^\ddag$  Tainan Hydraulics Laboratory, the National Cheng Kung University, Tainan, Taiwan}

   \begin{abstract}
A novel model of discretized energy cascade generated by Benjamin-Feir instability is presented. Conditions for  appearance of direct and inverse cascades are given explicitly, as well as conditions for stabilization of the wave system due to cascade termination. These results can be used directly for explanation of  available results of laboratory experiments and as basic forecast scenarios for planned experiments, depending on the frequency of an initially excited mode and  steepness of its amplitude.
\end{abstract}

{PACS: 05.60.-k, 47.85.-g, 47.20.-k}

\maketitle

\section{1. Introduction}

 Benjamin-Feir instability (BF-instability) is one of  fundamental principles of nonlinear water wave dynamics, \cite{BF67}. This phenomenon is of the utmost importance for  description of  dynamics and downshifting of energy spectrum among  sea surface waves, formation of freak (or giant) waves in  oceans and wave breaking. BF-instability and its physical applications have been profoundly studied during the last few decades, \cite{BF-all,BF-1,BF-2,TW99}, and its main features can be briefly summarized as follows:

\textbf{I.} initial exponential growth of the main side bands of the carrier wave;

\textbf{II.} late asymmetry of sidebands and temporary frequency downshifting;

\textbf{III.} discretized spreading of energy to higher and lower frequencies;

\textbf{IV.} existence or absence of near recurrence Fermi-Pasta-Ulam (FPU) phenomenon in no breaking regime of wave propagation, in  experiments with different parameters of initial excitation.

As for surface gravity waves on deep water resonant interactions occur at the third order,  BF-instability can be described at early stages of the process as interaction of three monochromatic wave trains: carrier ($\o_c$ ), upper ($\o_+=\o_c + \Delta \o$) and lower ($\o_-=\o_c - \Delta \o$) side-band waves with $\Delta \o >0$ which form a resonant quartet for one particular configuration which occurs when two of the waves coincide, with frequency resonance condition
\be \label{Phil}
\o_+ + \o_- = 2\o_c.
\ee
where $\o_i\equiv \o(\mathbf{k}_i)$ and $\mathbf{k}_i$ are notations for dispersion function and wave vector correspondingly.

In this Letter we present a novel model of BF-instability basing on the model of dynamical cascade generation in a wave system with narrow initial excitation, first introduced in \cite{CUP}. We  demonstrate that in the frame of our model a) main features \textbf{I.}-\textbf{IV.} of BF-instability are naturally reproduced; and b) dependence of the cascade form on details of initial excitation (choice of frequency and wave steepness) is in accordance with available experimental data.

 \section{2. Dynamical equations}

 Dynamical system corresponding to an isolated quartet
 reads:
 \bea \label{4primA}
 \begin{cases}
  i\, \dot{A}_1=  T A_2^*A_3A_4 +(\tilde \omega_1 - \o_1) A_1\,,\\
 i\, \dot{A}_2=T A_1^*A_3A_4 +(\tilde \omega_2 - \o_2) A_2\,, \\
 i\, \dot{A}_3=  T^* A_4^*A_1A_2 +(\tilde \omega_3 - \o_3) A_3\,, \\
 i\, \dot{A}_4=  T^* A_3^*A_1A_2 + (\tilde \omega_4 - \o_4)A_4\,, \\
\tilde \omega_j - \o_j = \sum_{i=1}^4(T_{ij}|A_j|^2 -   \frac12\,  T_{jj}|A_i|^2)\,,
 \end{cases}
  \eea  
where   interaction coefficients $T_{ij}= T_{ji}\= T_{ij}^{ij}$  and $T=T^{12}_{34} $
are  responsible for the nonlinear shifts of frequency and the energy exchange within a quartet correspondingly;
 $ (\tilde \omega_j - \o_j) $
are  Stokes-corrected frequencies and $A_j$ are slowly changing amplitudes of resonant modes in canonical variables. Explicit form of interaction coefficients for surface gravity waves is given in \cite{Kr94}.
Analytical solution of (\ref{4primA}) in terms of elliptic functions is found  in \cite{SS05} and is studied numerically.

Aiming to analyze quartet dynamics \emph{qualitatively} in the case when only one or two modes are initially excited, one can use
standard change of variables $A_j\= -i C_j \exp [-i \varphi_j]\,,$ and rewrite (\ref{4primA}) in amplitude-phase presentation as
\bea \label{4-prim}
\frac{d\, C_1 ^2}{dt}&=&\frac{d\, C_2 ^2}{dt} = -\frac{d\, C_3 ^2}{dt}  = -\frac{d\, C_4 ^2}{dt} \nn \\
&=&  2 |T| C_1C_2C_3 C_4\, \sin (\arg T -\varphi_{12,34})\,,
\eea
where  the dynamical phase
$
 \varphi_{12,34} \= \varphi_1+\varphi_2-\varphi_3-\varphi_4
$
corresponds to the chosen resonance conditions  of the form
 \be \label{res4}
 \o_1+\o_2=\o_3+\o_4, \ \mathbf{k}_1+\mathbf{k}_2=\mathbf{k}_3+\mathbf{k}_4.
 \ee
Sys.(\ref{4-prim}) has three independent  Manley-Rowe constants of motion
\bea \label{4MR}
I_{13}= C_1 ^2+ C_3 ^2\,, I_{14}= C_1 ^2+ C_4 ^2\,, I_{23}= C_2 ^2+ C_3 ^2.
\eea
Any linear combinations of these three are also constants of motion, e.g.
\bea\label{4MRa}
I_{24}&=& C_2 ^2+ C_4 ^2\,,\quad  I_{1234}= C_1 ^2+ C_2 ^2+C_3 ^2+ C_4 ^2 \,,
 \\  I_{1,2}&=&  C_1 ^2- C_2 ^2\,, \quad I_{3,4}= C_3 ^2- C_4 ^2\
\eea
and can be used for qualitative analysis of a quartet dynamics for specific initial conditions.

\emph{Case 1: one mode is initially excited.} Let initially the mode $\o_1$ be excited, i.e. at the time moment $t=0$ we have
\label{4ic} \be\label{4icA} C_{1,0}\gg C_{2,0}\simeq C_{3,0}\simeq C_{4,0}\= c_0\,,
\ee
where $C_{j,0}\=C_j(t=0)$. It follows from (\ref{4MR}), that the amplitudes $C_2(t)$,   $C_3(t)$ and  $C_4(t)$, being initially small, remains small at all time moments $t>0$. Indeed,
\bea
2c_0^2 \sim I_{24}+I_{14}-I_{3,4}=C_2 ^2+C_3 ^2 \nn \\
 \RA C_2(t) \sim C_3(t) \sim c_0,\\
0 \sim I_{13}-I_{14}=C_3 ^2 -C_4 ^2 \RA  C_4(t) \sim c_0,
\eea
for any $t>0,$ i.e. the energy transfer in this case is essentially suppressed.

\emph{Case 2: two modes are initially excited.} There exist two different types of modes' pairs which should be regarded separately, \cite{CUP}:
 one-side-pair
and two-side-pair, referring to  resonance conditions of the form (\ref{res4}).   Accordingly, there are two 1-pairs of modes' frequencies:
 \be
 \mbox{\emph{1-pairs}:}\quad (\o_1,\o_2)\,, \ (\o_3,\o_4)\,,  \ee
 and four 2-pairs of modes' frequencies:
\be \mbox{\emph{2-pairs}:}\quad (\o_1,\o_3)\,, \ (\o_1,\o_4)\,, \ (\o_2,\o_3)\,, \ (\o_4,\o_3)\,.
 \ee 
\emph{Case 2a: 2-pair is initially excited.} Let
\be\label{4icPa} C_{1,0}\simeq  C_{3,0}\gg  C_{2,0}\simeq C_{4,0}\= c_0\,,
\ee
then again it follows from the form of Manley-Rowe constants of motion that
$
C_2(t) \sim C_4(t) \sim c_0.
$

\emph{Case 2b: 1-pair is initially excited.} Regard 1-pair ($\o_1\,,\o_2$) with initial modes' amplitudes as follows:
 \be\label{4icPb} C_{1,0}\sim   C_{2,0}\gg  C_{3,0}\simeq C_{4,0}\ .
\ee
There exists no restriction on modes' growth originating from the Manley-Rowe constants of motion, and resulting evolution depends on the details of the initial energy distribution within a quartet.

 During initial evolution, during which inequalities~(\ref{4icPb}) still hold, one can neglect the effect of modes  $\o_3, \o_4$ on modes $\o_1,\o_2$ and \Eqs{4primA} can be solved explicitly:
 \bea \label{sol}
 \begin{cases}
  A_1(t)=C_{1,0}\exp (i \Delta_{1,0}t),\\
 A_2(t)=C_{2,0}\exp (i \Delta_{2,0}t),\\
 A_3(t)= C_{3,0}\exp \big[ (i\, \tilde{\o}_3 + \nu_{12} \big ) t \big]\,, \\
   A_4^*(t)= C_{4,0}\exp \big[ (- i\, \tilde{\o}_4 + \nu_{12} \big ) t \big]\,,
\end{cases}
\eea
where
\bea
\Delta_{1,0} &=& \frac{T_{11}} 2C_{1,0}^2+ T_{12}C_{2,0}^2\, \nn \\
\Delta_{2,0} &=& \frac{T_{22}} 2C_{2,0}^2+ T_{12}C_{1,0}^2,\nn \\
 \Delta_{3,0} &=&  T_{31}  C_{1,0}^2+ T_{32}C_{2,0}^2\,, \nn \\
 \Delta_{4,0}  &=&   T_{41}  C_{1,0}^2+ T_{42}C_{2,0}^2\,, \nn \\
  \tilde{\o}_3 &=& (\Delta_{1,0}+\Delta_{2,0}-\Delta_{3,0}+\Delta_{4,0} )/2\,, \nn \\
  \tilde{\o}_4 &=& (\Delta_{1,0}+\Delta_{2,0}+\Delta_{3,0}-\Delta_{4,0} )/2\,, \nn
\eea
\bea
    \nu_{12} ^2 &=& |\C P|^2- \frac 14 \Big( \sum_{j=1}^4 \Delta_{j,0}\Big)^2 \label{nu} \\
     &=& |T|^2 C_{1,0}^2 C_{2,0}^2-   \big(\C T _1  C_{1,0}^2+ \C T _2  C_{2,0}^2\big )^2\big / 4\, , \nn \\
\C P &=& T^*  C_{1,0} C_{2,0} \exp [\, i\, (\Delta_{1,0}+\Delta_{2,0})\, t\, ]\,, \nn \\
 \C T _1&=& \frac 12 T_{11}+ T_{12}+T_{13}+ T_{14}\,, \nn \\
  \C T _2&=& T_{12} + \frac 12 T_{22}+ T_{23}+T_{24}\,. \nn
\eea
Accordingly, the evolution of the amplitudes $A_3$ and $A_4$ is defined by the sign of the increment $\nu_{12} ^2$ given by (\ref{nu}).

If $\nu_{12} ^2> 0$,  (\ref{nu}) predicts exponential grow of amplitudes $A_3$ and $A_4$. In this case energy goes from the initially excited 1-pair $(\o_1\,, \o_2)$  to the second  1-pair $(\o_3\,, \o_4)$ with  characteristic time $\simeq 1/\nu_{12}$. Similarly, if the 1-pair $(\o_3\,, \o_4)$ is initially excited, energy can effectively  go to the 1-pair $(\o_1\,, \o_2)$ , if $\nu_{34}>0$.

However, (\ref{nu}) does not guarantee that $\nu_{12}^2$ is positive.  For instance, if $C_{1,0}\gg C_{2,0}$ or $C_{1,0}\ll C_{2,0}$,  $\nu_{12}^2$ is negative for any interaction coefficient while in the case  $C_{1,0}= C_{2,0}$ the sign of $\nu_{12}^2$ depends on the relations between $|T|$ and $(\C T_1+\C T_2)/2$:
  \be  \label{4instE}
  \nu_{12}^2 =  \big[ |T|^2  -   \big(\C T _1  + \C T _2\big )^2/4\big] C_{1,0}^4\,,
  \ee

If $\nu_{12}^2 <0$,  the solution (\ref{sol}) yields  pure oscillatory behavior of the amplitudes  $A_3, A_4$   with  frequencies $\tilde{\o_{3}}\pm |\nu_{12}|$   and $\tilde{\o_{4}}\pm |\nu_{12}|$ correspondingly.

\section{3. Generation of a cascade}

The general model of dynamical cascade generation in 3- and 4-wave systems with narrow initial excitation is sketched in \cite{CUP}. In this Letter  we work out the details and apply it for the description of BF-instability.

\emph{At the initial step, $n=0$,} frequency resonance conditions have the same form  (\ref{Phil}) for both direct and inverse cascade
while the only form of a  quartet in which \emph{one-mode excitation} yields generation of resonant interactions  reads
  \be \label{Has}\o_1+\o_2=2\o_3\ee
and only in the case when the mode with frequency $\o_3$ is excited. This occurs due to Hasselmann's criterion of instability for 4-wave systems, \cite{Has67}.
For all other configurations and initial excitations, any single mode in a quartet is neutrally stable. The most effective resonance takes place if (\ref{Has}) is satisfied exactly,
i.e. $\Delta \o$ should be the same for both $\o_+ $ and $\o_-$.

Hasselmann's criterion can be applied at each further step yielding the general form (\ref{Has}) at each step of a cascade. To simplify further presentation we introduce notation $\o_{\pm n}=\o_c \pm n \Delta \o$.

\emph{At the step $1$}, a couple of new modes $\o_1 $ and $\o_{-1}$ is generated. From pure kinematical considerations, at the next steps of a cascade all possible quartets of the form
\be
\o_c \pm \o_1 \mp \o_{-1} = \o_c \pm 2\Delta \o
\ee
are possible. However,  different dynamical properties of 1-pair and 2-pair in a quartet define  two possible combinations only, yielding exact frequency resonance condition for quartets of the form (\ref{Has}) \emph{with excited 1-pair of modes}:
\bea \label{st1-both}
\o_c  + \o_c  + 2\Delta \o = 2(\o_c + \Delta \o) \nn \\
\quad \quad  \quad  \RA \o_{c}+\o_{2} = 2\o_{1},\label{st1a}\\
\o_c  + \o_c  - 2\Delta \o = 2(\o_c - \Delta \o)\nn \\
\quad \quad \quad  \RA  \o_{c}+\o_{-2} = 2\o_{-1}\label{st1b}.
\eea
Accordingly, the beginning of direct and inverse cascades is given by (\ref{st1a}) and (\ref{st1b}).
As $\o_{c}+\o_{\pm 2} = 2\o_{\pm 1},$ frequency shift $|\o_{c}-\o_{\pm 2}| = 2 \Delta \o$ occurs (its sign is opposite for direct and inverse cascade).

Similarly, \emph{at the step $n$} we have
\bea \label{stn-both}
\o_c  + \o_c  + 2n\Delta \o = 2(\o_c + n\Delta \o)\nn \\
\quad \quad  \quad  \RA \o_{n+1}+\o_{n-1} = 2\o_{n}\label{stna}\\
\o_c  + \o_c  - 2n\Delta \o = 2(\o_c - n\Delta \o)\nn \\
\quad \quad  \quad  \RA \o_{-(n+1)}+\o_{-(n-1)} = 2\o_{-n}\label{stnb},
\eea
and the
frequency shift is again $2 \Delta \o$, i.e. its magnitude is the same at each step of both direct and inverse cascade. The complete system describing both direct and inverse cascade takes now the following form:
\bea \label{cas-gen}
\begin{cases}
\o_+ +\o_- =2\o_c \\
\o_c+\o_{\pm 2} = 2\o_{\pm 1}\\
\o_{\pm 1}+\o_{\pm 3} = 2\o_{\pm 2}\\
...\\
\o_{\pm (n+1)}+\o_{\pm (n-1)} = 2\o_{\pm n}.
\end{cases}
\eea
Direction of  cascade depends on the sign chosen in the lower index  $\o_{\pm n}$. Indeed, the
choice of $\o_{-n}$ generates a sequence of frequencies $\o_c > \o_{-1} > ... > \o_{- n},$ i.e. inverse cascade, while the
choice of $\o_{+n}$ yields direct cascade $\o_c < \o_{1} < ... < \o_{n}.$

\section{4. Dynamics of cascade}

Exponential growth of the two main side-bands $\o_{1},\o_{-1}$  at the initial step $n=0$ is caused by the resonance (\ref{Has}) at the expense of the initially excited mode $\o_{3}=\o_{c}$ and has non-dimensional increment (in physical variables, i.e. $I \sim \nu_{12}$)
\be \label{BFI-incr}
0< I=\eps^2\sigma(2-\sigma^2)^{1/2}/2 < 1,
\ee
where $\sigma=\Delta \o /\eps \o_c$ is the ratio of the relative resonance frequency bandwidth of the wave train to the initially excited wave steepness $\eps=C_ck_c$.

Maximal growth rate is the same for sub- and super-harmonics of the excited wave:
\be \label{BFI-incr-max}
I_{max}=\eps^2 / 2 = C_c^2 {k_c}^2 / 2
\ee
At the next step $n=1$ direct and inverse cascade  will be be initialized by two already exciting modes $\o_1$  and $\o_{-1}$  correspondingly.

However, the important observation is that at the first step of a cascade the maximum of the increments for direct $I_{\mathbf{dir},+}$ and inverse $I_{\mathbf{inv}, -}$  cascades differ:
\be
I_{\mathbf{dir},+} \sim C_+^2 k_+^2/2 \quad \mbox \quad I_{\mathbf{inv}, -} \sim C_-^2 k_-^2/2.
\ee
This means that in the regime of symmetrical growth of the main side-bands $C_+ \sim C_-$, the \emph{asymmetrical behavior of the growing modes} will be observed since
\be
k_+>k_- \quad \RA \quad I_{\mathbf{dir},+} > I_{\mathbf{inv}, -}.
\ee
Similar considerations allow us to conclude that
 at the \emph{n}-th step of the cascade again
\be
 I_{\mathbf{dir},+n} > I_{\mathbf{inv}, -n}.
\ee
 The last equation shows that at each step, increment of instability for  direct cascade is bigger than increment of instability for  inverse cascade.

This clearly explains \emph{the frequency downshift phenomenon} -- the main higher side-band component spreads energy to higher frequency modes
\be \o_c < \o_{1} < ... < \o_{n}\nn \ee
by direct cascade mechanism much faster than  the main lower side-band by the inverse cascade with corresponding
sequence of frequencies
\be \o_c > \o_{-1} > ... > \o_{- n}. \nn \ee
Accordingly, main lower side-band mode finally will be dominant which is manifestation of frequency downshift phenomenon.

 This scenario can be clearly observed e.g. in \cite{TW99} where laboratory observations of wave group evolution, including breaking effects are presented. In particular, it is stated there (p. 223) that "the initial distribution of energy (...) is altered by the transfer of energy to free waves $\o_0\pm \delta\o$, first noticeably to $n=+2,$ then $n=-2,+3,$ etc. These relatively fast transfers seems to be a consequence of detuned resonances of which first is
\bea \label{WT-first-step}
\begin{cases}
(\o_0)\mp(\o_0-\delta\o)\pm(\o_0+\delta\o)=\o_0\pm2\delta\o,\\
(k_0)\mp(k_0-\delta k)\pm(k_0+\delta k)=k_0\pm2\delta k +\Delta k,
\end{cases}
\eea
in which $\Delta k $ is a small detuning factor." As in our notations $\o_0=\o_c$ and $\delta\o=\Delta \o$, the frequency resonance condition above is just a more compact form of two frequency resonance conditions (\ref{st1a}),(\ref{st1b}) for quartets of the form (\ref{Has}) with excited 1-pair of modes.

Experiments presented in \cite{TW99} have been performed in a large wave tank (50 x 4.2 x 2.1 m), for the range of carrier wave initial steepness  $0.12 < \eps < 0.28$ and  wavelengths 1 to 4 m. Both direct and inverse discretized energy spectra have been clearly observed (see e.g. Fig.20 from \cite{TW99}),  breaking and recurrence phenomena have been reproduced.

The authors concluded that
"a serious challenge is imposed by these results to any predictive method
of ocean wave evolution in which downshifting depends solely on slow, high-order,
resonant wave interactions" which
"might very well dominate the downshifting of energetic waves in the
ocean, too" (\cite{TW99}, p.226).

We conclude here that Tulin and Waseda in \cite{TW99} came very close to the conclusive solution of this problem while describing the first few steps of dynamical cascade. Only one final step -- realization that corresponding detuned resonances  build  clusters of the form (\ref{cas-gen}) -- has not been done.

\section{5. Termination of  cascade}

Energy cascading will be terminated as soon as the BF-criterion of instability (\ref{BFI-incr}) is violated, i.e. if $I \approx 0$ (stabilization of the wave system due to transition to linear regime) and if $I>1$ (substantial increasing of the cascading modes' steepness and transition to breaking regime). In the last case
\be \label{BFI-incr1}
\sigma^2=(\Delta \o/\o_c/\eps)^2 >2.
\ee
For the $n$-th step of  cascade, the magnitude of $\sigma^2$  can be estimated as follows:
\be \label{star}
\sigma^2 = \frac{(\Delta \o)^2}{(\o_n \eps)^2} = \frac{(\Delta \o)^2}{(\o_c + n\Delta \o)^2 C_n^2(k_c+n\Delta k)^2} >2.
\ee
We can obtain upper estimate of the energy spreading at the $n$-th step of the cascade in the resonance of form (\ref{Has}) by its maximal value, when the excited carrier wave energy will be totally and equally distributed between two resonance side-band modes:
\be
C_n^2=C_{n-1}^2+C_{n+1}^2 \sim 2 C_{n+1}^2
 \RA  C_{n+1}^2  \le C_n^2/2. \nonumber
\ee
This means that \emph{the general upper estimate} for the $n$-th step of cascade has the form
\be \label{sq_amp}
C_n^2 \le C_0^2/2^n.
\ee
After substituting (\ref{sq_amp}) to the condition (\ref{star}), this yields
\be \label{2stars+}
\sigma^2 = \frac{(\Delta \o)^2}{(\o_n \eps)^2} = \frac{2^n (\Delta \o)^2}{(\o_c + n\Delta \o)^2C_0^2(k_c+n\Delta k)^2} >2.
\ee
The latter inequality shows clearly  that the stability criterion  will be reached  after \emph{a finite number of steps} for the direct energy cascade; correspondingly, energy spreading to higher frequencies will be terminated due to the growth of nonlinearity and consequent breaking effects which is in accordance with laboratory results \cite{TW99}.

For the inverse cascade we can estimate the stabilization condition in a similar manner as
\be \label{2stars-}
\sigma^2 = \frac{(\Delta \o)^2}{(\o_{-n} \eps)^2} = \frac{2^n (\Delta \o)^2}{(\o_c - n\Delta \o)^2 C_0^2(k_c-n\Delta k)^2} >2,
\ee
i.e. the inverse cascade will be terminated even faster than the direct cascade.

These estimates have been made in accordance with classical results  (\ref{BFI-incr}),(\ref{BFI-incr-max}) and are valid only for small enough initial steepness of the excited wave $\eps \sim 0.1.$

For moderate and high initial wave train steepness $\eps \sim 0.1 \div 0.4$,   improved results for the growth of  increments and a criterion of instability have been obtained by Dysthe, \cite{DY79}.
In this case, instability violates if
\be
\sigma^2>2\Big( 1-\frac{2\Delta \o}{\o_c}  \Big)
\ee
This obviously means that termination of energy cascades can happen even earlier for a higher initial steepness of carrier wave.

 As it is shown in  \cite{DY79},
maximal growth rate is modified by the following way (see  Eq.(3.10), p. 111, notations as in the cited paper):
\be
\g_m=\frac{1}{2}(1-2A_0)A_0^2 \nn
\ee
where $A_0=ak$ is steepness of carrier wave. Accordingly, in our notations
\be \label{inc-dys}
I_{max}  = \frac{1}{2}\eps^2 (1 - 2\eps)=
\frac{1}{2}C_c^2 k_c^2 (1- 2 C_c k_c ).
\ee
Expressions for $ I_{\mathbf{dir},+n}$ and $ I_{\mathbf{inv}, -n}$ at the first cascade's step read
\bea
I_{\mathbf{dir},+}=\frac{1}{2}C_+^2 k_+^2 (1- 2 C_+ k_+), \quad k_+=k_c+\Delta k,\\
I_{\mathbf{dir},-}=\frac{1}{2}C_-^2 k_-^2 (1- 2 C_- k_-), \quad k_-=k_c-\Delta k,
\eea
and in the regime of symmetrical growth of the main side-bands $C_+ \sim C_- \sim C $ we have
\bea \label{incr-diff}
I_{\mathbf{dir},+}-I_{\mathbf{dir},-}\sim  \nn \\
\frac{1}{2}C^2 [k_+^2 (1- 2 C k_+)-k_-^2 (1- 2 Ck_-)]  \nn \\
= \frac{1}{2}C^2[(k_+^2-k_-^2) -2C(k_+^3-k_-^3)]  \nn \\
=\frac{1}{2}C^2(k_+-k_-)[k_+ + k_- -2C(k_+^2+k_+k_-+k_-^2)]  \nn \\
=2C^2\Delta k [k_c - C(3k_c^2+(\Delta k)^2)].
\eea
As $C, k, \Delta k$ are positive, it follows from (\ref{incr-diff}) that $I_{\mathbf{dir},+}>I_{\mathbf{dir},-}$ only if the following condition holds
\be
k_c - C(3k_c^2+(\Delta k)^2) >0,
\ee
which is satisfied in the finite range of wavelengths $k_c.$ Similar considerations show that at $n$-th cascade's step, $I_{\mathbf{dir},+n}>I_{\mathbf{dir},-n}$ is also satisfied in the finite range of $k_c$ such that $k_c - C(3k_c^2+(n\Delta k)^2) >0$. For all other $k_c$, maximal increment for inverse cascade will be larger than for direct cascade,  $ I_{\mathbf{inv}, -n}> I_{\mathbf{dir},+n}.$

Thus, by considering  two forms of  instability increment given by (\ref{BFI-incr-max})  and (\ref{inc-dys})
for the energy cascade concept, we can make the conclusion that for essentially high initial steepness $\eps \gtrsim 0.25$  the inverse dynamical energy cascade to lower frequencies may dominate. In general, both directions of cascade may be significant, unlike the case of a small initial steepness of the excited wave.

\section{6. Conclusions}

\noindent\textbullet~The main features \textbf{I.}-\textbf{IV.} formulated in the Introduction have been previously studied mainly numerically, while our model gives a clear explanation of \emph{the physical origin of the observed phenomena}.

\noindent\textbullet~Moreover, several fundamental facets of BF-instability
that have not been investigated theoretically until now -- cascade direction, its dependence on the initial conditions and finiteness of the number of cascading modes -- are adequately and constructively described by our model:

\textbullet\textbullet~Direction of  cascade depends on the initial wave steepness of the wave train; for small enough steepness $\eps \sim 0.1$ direct energy cascade prevails, while for  high enough steepness $\eps \gtrsim 0.25$ inverse cascade and energy spreading to lower frequency modes may be comparable with direct cascade rates.

\textbullet\textbullet~An increase of initial steepness  may reduce the number of steps for energy cascades in both directions.

\textbullet\textbullet~Frequency downshifting in the wave system is caused by the direct energy cascade and spreading of energy to higher frequencies at the expense of the main super side-band mode; this process dominates for  small to moderate initial steepness of the excited waves.

\textbullet\textbullet~Inverse cascade can be expected to be the prevailing process for high enough basic carrier wave steepness.

\textbullet\textbullet~Only a finite number of steps in both direct and inverse energy cascade leading to both breaking or stabilization regimes satisfy the BF-criterion.

 \textbullet\textbullet~ Fermi-Pasta-Ulam recurrent process may occur if at some cascade step modes are generated which are in \emph{exact} resonance, i.e.  $\Delta k=0$ in (\ref{WT-first-step}) (this is only \emph{necessary condition}). This general prediction is in accordance with experimental results  where it was shown that the observed recurrence is very close to the NLS or three-wave system of the form (\ref{Has}) and
"neither a reduction in the number of waves per group nor downshifting
of the spectral energy" is observed in the experiments of Tulin and Waseda (\cite{TW99}, p. 210) conducted in the laboratory tank of the sizes 50 x 4.2 x 2.1 m.

The evolution of wave trains on an effectively much longer fetch than previous studies (330 x 5 x 5 m) was experimentally investigated in \cite{BF-1,BF-2} with following results.
A very long scaled wave modulation with several modulation loops demonstrates re-stabilization processes: finally the periodic modulations of wave train are observed at the latest stage with the most energetic lower side-band wave. This means the final termination of the wave cascade mechanism and stabilization of the system with essentially large steepness in accordance with our model.

For small initial steepness of the wave packet $\eps \sim 0.1$ Benjamin-Feir instability leads to growing of the main pair side-band modes. Asymmetrical growth of side-bands with prevailing of the lower side-band mode and discretized energy cascade to the  higher frequencies  was also observed in the experiments \cite{BF-1,BF-2}. Near recurrence FPU phenomenon was clearly seen at the latest stages of the wave propagation: most part of wave energy revert back to the carrier frequency mode. For large enough initial steepness of the wave packet $\eps \sim 0.15$ to 0.25 wave steepness during propagation leads to wave breaking. Periodic modulation and demodulation of wave trains are found at post-breaking stage, in which the energy of wave train transfers between the carrier wave and a pair of sidebands. Only partial FPU phenomenon was observed in this case - essential part of energy is lost due to breaking.

Necessary and sufficient conditions for  manifestation of this phenomenon  depend explicitly on the relation between wavelength and initial phase of the carrier wave and the aspect ratio of the laboratory tank (see (\ref{4instE}) and remark afterward;  more details can be found in \cite{CUP}).
Deduction of their  explicit form  is outside the scope of this Letter where we concentrated on the study of non-periodic energy transfer due to the novel mechanism - dynamic cascade given by (\ref{cas-gen}). 

\textbullet~Our model of BF-instability can be further refined in many aspects. For instance, dissipation can  easily be included at each step by changing of $C_n$ to $p_n C_n$ in (\ref{BFI-incr1}), with some constants $p_n, \ 0<p_n<1$. Accordingly, the form of the increments  $I_{\mathbf{dir},+n}, I_{\mathbf{inv}, -n}$ will be modified and the number of cascade's steps will be reduced. Detailed study of the effect of dissipation can be found in \cite{Os10}.

\textbullet~This model is quite general and describes basic energy cascade in \emph{an arbitrary 4-wave system with narrow initial excitation} though the estimate (\ref{BFI-incr}) and definition of the small parameter $\eps$ might change, depending on the specifics of the wave system. In particular, the model can be used directly as a basic description of wind generated instabilities of surface water waves from which extreme (or rogue) waves originate, e.g. see \cite{KS08}  for numerical simulations and \cite{freak-lab1,freak-lab2} for laboratory study. Theoretical study of the  role of Benjamin-Feir instability in formation of extreme waves can be found e.g. in \cite{freak-theor}.\\

 \noindent
{\textbf{Acknowledgements.}} Authors are grateful to  anonymous Referee for valuable remarks and suggestions. E.K. acknowledges the
support of the Austrian Science Foundation (FWF) under project
P22943-N18 "Nonlinear resonances of water waves". I.Sh. acknowledges Russian Foundation for Basic Research (10-02-92005 HHC-a).


\begin{thebibliography}{99}
\bibitem{BF67}
 Benjamin, T. B., and J. E. Feir.
     \newblock \emph{Fluid Mech.}, \textbf{27}: 417 (1967).

\bibitem{BF-all} Lo, E., and C.C. Mei. \emph{Fluid Mech.} \textbf{150}: 395 (1985); Segur, H., D. Henderson, J. Hammack, C.-M. Li, D. Phei and K. Socha. \emph{Fluid Mech.} \textbf{539}: 229 (2005);   T. Bridges, F. Dias. \emph{Phys. Fluids} \textbf{19}: 104104 (2007);  and many others.

\bibitem{BF-1}
Hwung, H. H., Wen-Son Chiang, and Shih-Chun Hsiao.
\emph{Proc. R. Soc. A} \textbf{463}: 85 (2007).

\bibitem{BF-2} Hwung, H.-H., W.-S. Chiang, R.-Y. Yang and I. V. Shugan. \emph{Eur. J. Mechanics B/Fluids} \textbf{30}: 147 (2011).

\bibitem{TW99}
 Tulin, M. P., and  T. Waseda.
     \newblock \emph{Fluid Mech}. \textbf{378}: 197 (1999).

\bibitem{CUP} Kartashova, E.
\emph{Nonlinear Resonance Analysis}
     \newblock (Cambridge University Press, 2010).

\bibitem{SS05}
M. Stiassnie, and L. Shemer.
     \newblock \emph{Wave motion} \textbf{41}: 307 (2005).

 \bibitem{Has67}
K. Hasselmann.
     \newblock \emph{Fluid Mech.}, \textbf{30}: 737 (1967).

\bibitem{DY79}
Dysthe, K. B.
     \newblock \emph{Proc. R. Soc. A} \textbf{369}: 105 (1979).

\bibitem{CFGK06}
Clamond, D. , M. Francius, J. Grue, and C. Kharif. \emph{Eur. J.
Mech. B, Fluids} \textbf{25}: 536 (2006).

\bibitem{Os10}
Osborne, A. R.
     \emph{Nonlinear Ocean Waves and the Inverse Scattering Transform} (International Geophysics Series \textbf{97}, Academic Press, 2010).

\bibitem{KS08}
Kuznetsov, S., and Ya. Saprykina.
     \newblock Proc. of workshop "ROGUE WAVES 2008" (Brest, France, October 2008), p. 99.

\bibitem{freak-lab1}
Waseda, T., H. Tamura and T. Kinoshita.
     \newblock Proc. of workshop "ROGUE WAVES 2008" (Brest, France, October 2008), p. 207.

\bibitem{freak-lab2}
Slunyaev, A.,  A. Ezersky, D. Mouaz\'{e} and W. Chokchai.
     \newblock Proc. of workshop "ROGUE WAVES 2008" (Brest, France, October 2008), p. 209.

\bibitem{freak-theor}
Yuen, H. C, B. M. Lake.
\newblock \emph{Adv. App. Mech.} \textbf{22}: 67 (1987);
Infeld. E.,  and G. Rowlands.
\newblock  \emph{Nonlinear waves, solitons and chaos} (Cambridge University Press, 2000); Kharif, C., E. Pelinovsky and A. Slunyaev. \emph{Rogue waves in the ocean} (Springer, 2009).

\bibitem{Kr94}
V. P. Krasitskii.
    \newblock On reduced equations in the Hamiltonian theory of weakly non-linear surface waves.
    \newblock \emph{Fluid Mech. }{\bf 272} (1994), 1--20.
\end{thebibliography}
 \end{document}